\documentclass[a4,12pt]{article}
\usepackage{epsf}

\newcommand{\nn}{\nonumber\\}
\newcommand{\bra}[1]{\left<#1 \right|}
\newcommand{\ket}[1]{\left| #1 \right>}
\newcommand{\abs}[1]{\left| #1 \right|}

\newcommand{\Q}{Q_{\rm B}}
\newcommand{\half}{\frac{1}{2}}

\renewcommand{\thepage}{}
\makeatletter
\@addtoreset{equation}{section}
\renewcommand{\theequation}{\thesection.\@arabic\c@equation}
\makeatother

\begin{document}
\begin{titlepage}
\title{
\vspace*{-4ex}
\hfill{\normalsize hep-th/0302182}\\
\vspace{4ex}
\bf Tachyon Condensation and Universal\vspace{2ex}\\
Solutions in String Field Theory
\vspace{5ex}}
\author{Tomohiko {\sc Takahashi}\thanks{E-mail address:
 {tomo@asuka.phys.nara-wu.ac.jp}}
\vspace{3ex}\\
  {\it Department of Physics, Nara Women's University}\\
  {\it Nara 630-8506, Japan}}
\date{February, 2003}
\maketitle
\vspace{7ex}

\begin{abstract}
\normalsize
\baselineskip=19pt plus 0.2pt minus 0.1pt
We investigate a non-perturbative vacuum in open string field theory
expanded around the analytic classical solution which has been found
in the universal Fock space generated by matter Virasoro generators
and ghost oscillators. We carry out level-truncation analyses up to
level (6,\,18) in the theory around one-parameter families of the
solution. We observe that the absolute value of the vacuum energy
cancels the D-brane tension as the approximation level is increased,
but this non-perturbative vacuum disappears at the boundary of the
parameter space. These results provide strong evidence for the
conjecture that, although the universal solutions are pure gauge in
almost all the parameter space, they are regarded as the tachyon vacuum
solution at the boundary.  
\end{abstract}
\end{titlepage}

\renewcommand{\thepage}{\arabic{page}}
\setcounter{page}{1}
\baselineskip=19pt plus 0.2pt minus 0.1pt
\section{Introduction}

String field theory is a promising approach to investigate
non-perturbative aspects of  string theory. As conjectured 
by Sen \cite{rf:SenUniv,rf:Sen}, we can describe the annihilation
process of D-branes by using the condensation of the tachyon in open
string field theory, in which there is a stable vacuum
\cite{rf:KS-tachyon} and the energy difference between the stable and
unstable vacua is in precise agreement with a D-brane tension
\cite{rf:SZ-tachyon,rf:MT,rf:GR}. Since these discoveries much progress
has been made in string field theory, particularly in the formulation of
vacuum string field theory \cite{rf:RSZ-VSFT}, where some exact results
are obtained \cite{rf:VSFT}.
However, despite these developments, we have not yet found the analytic
classical solution, which is eagerly awaited, corresponding to the
tachyon vacuum in open string field theory. 

While there are some formal attempts to construct analytic
solutions in string field theory
\cite{rf:TT,rf:TT2,rf:KishiOhmo,rf:Kluson},
strange problems often arise from such formal solutions.
For example, if $Q_{\rm L}$ stands for the left-half integration of the
BRS current and 
$I$ is the identity string field,
$Q_{\rm L}I$ is a formal solution in purely cubic string field theory
\cite{rf:HLR}.  
The equation of motion is given by $\Psi*\Psi=0$
and then $-Q_{\rm L}I$ is also a solution, around which the kinetic
operator becomes $-\Q$ and the theory should describe open strings with a
negative tension. Therefore ordinary D-branes and
negative tension branes are realized with the same energy density.
We find another example in the context of vacuum
string field theory as pointed out in \cite{rf:RSZ-VSFT}.
From similar discussions, it follows that the solutions leading to pure
ghost kinetic operators provide the same energy density as the
perturbative vacuum. Usually, these unreliable results
are caused by midpoint singularities in a half string formulation. In
the former example, though the operator $Q_{\rm L}^2$ appears in solving
the equation of motion, this operator itself is
ill-defined due to a midpoint singularity, 
as pointed out for example in \cite{rf:TT2}. We find the same
singularity in the latter case. 

Fortunately, the analytic solutions found in \cite{rf:TT,rf:TT2} escape
from all these problems related to the midpoint singularity.
In addition they have many remarkable features:
The solutions can be expressed by
states in the universal Fock space which is spanned by matter Virasoro 
generators and ghost oscillators acting on the $SL(2,R)$ invariant
vacuum (so we call them {\it universal solutions}). 
This universality is necessary for the solution corresponding to the 
tachyon vacuum \cite{rf:SenUniv}. 
Secondly, open string excitations disappear after the string field
condensation to a specific class of the universal solutions
\cite{rf:KT}. This property is also
indispensable for the tachyon vacuum solution.
Consequently, we naturally expect that a certain kind of the universal
solutions is regarded as the tachyon vacuum solution.

Even if the universal solutions are irrelevant to the tachyon vacuum,
it should be emphasized that
they have intimate relations to gauge symmetry in string field theory,
which is an underlying principle in the theory
and which is much larger symmetry than existing in the low energy
effective theory. 
We can construct the universal solutions with a parameter.
They are pure gauge solutions in almost all the region of the parameter,
but they become non-trivial solutions at the boundary of the parameter
space. Hence, the non-trivial universal solutions can be
regarded as a kind of singular gauge transformations from the
perturbative vacuum \cite{rf:TT2}.
Moreover, 
the gauge symmetry and the annihilation mechanism
of open strings are inseparable.
In the theory around the non-trivial solution, the
kinetic operator is given by the modified BRS charge which has the
cohomology with different ghost numbers from the original
cohomology. Therefore
all of on-shell states are reduced to gauge degrees of freedom
in the gauge unfixed theory and then open string
excitations disappear after the condensation \cite{rf:KT}.

Although the tachyon vacuum solution has already been obtained
approximately in the level truncated theory, 
we can not so simply compared the universal solutions with the
level truncated solution,
because the gauges of these solutions differ.
The energy density of the universal solutions, then, should 
be calculated in order to clarify the relation between these
solutions. However, if we try to calculate the potential height
directly, there are some technical problems which are remained to be
resolved.  To avoid these we should adopt other approaches at
present. 

The purpose of this paper is to apply level truncation scheme
to investigate the theory around the universal solutions, and to
determine the potential height of the solutions indirectly.
If a one-parameter family of the solutions can be interpreted as
explained above, the theory around the solutions should
describe the perturbative vacuum in the almost region of the parameter
and the tachyon vacuum at the boundary. Hence, we should 
observe the situation that in the theory for the almost parameter,
there is a non-perturbative vacuum which gives the same energy
density as the D-brane tension, but the non-trivial vacuum disappears at
the endpoint. 

In Section 2 we review the universal solutions in string field
theory with some new results,
and we explain the difficulty of calculating the potential height. In
Section 3 we analyze the non-perturbative vacuum in the theory around the
universal solutions up to level (6,18). Our results strongly suggest
that the non-trivial universal solution corresponds to the tachyon
vacuum. In Section 4, we give summary and discussions.

\section{Classical solutions and potential heights}
\subsection{universal solutions}

The action of cubic open string field theory is given by
\cite{rf:CSFT}
\begin{eqnarray}
\label{Eq:action}
 S=-\frac{1}{g^2}\int \left(\frac{1}{2}\Psi*\Q\Psi
+\frac{1}{3}\Psi*\Psi*\Psi\right).
\end{eqnarray}
By variation of the action, we find the classical equation of motion
\begin{eqnarray}
\label{Eq:eqm}
 \Q\Psi+\Psi*\Psi=0.
\end{eqnarray}
One of the analytic solutions with translational invariance has been
found as \cite{rf:TT2} 
\begin{eqnarray}
\label{Eq:solution}
 \Psi_0 = Q_{\rm L}(e^h-1)I-C_{\rm L}((\partial h)^2 e^h)I,
\end{eqnarray}
where $I$ denotes the identity string field.
The operators $Q_{\rm L}$ and 
$C_{\rm L}$ are defined by 
\begin{eqnarray}
\label{Eq:QCdef}
 Q_{\rm L}(f)=\int_{C_{\rm left}} \frac{dw}{2\pi i}f(w)J_{\rm B}(w),
\ \ \  C_{\rm L}(f)=\int_{C_{\rm left}} \frac{dw}{2\pi i}f(w) c(w),
\end{eqnarray}
where $J_{\rm B}(w)$ and $c(w)$ are the BRS current and the ghost field,
respectively, and $C_{\rm left}$ stands for the contour along the left-half
of strings. 
The function $h(w)$ satisfies $h(-1/w)=h(w)$ and $h(\pm
i)=0$.

The solution (\ref{Eq:solution}) obeys the equation
of motion (\ref{Eq:eqm}) in the following.
The anti-commutation relations of $Q_{\rm L}$ and $C_{\rm L}$
are given by
\begin{eqnarray}
\label{Eq:QC1}
 \{Q_{\rm L}(e^h-1),\,Q_{\rm L}(e^h-1)\}&=&
    2\{\Q,\,C_{\rm L}((\partial h)^2 e^{2h})\},\nn
 \{Q_{\rm L}(e^h-1),\,C_{\rm L}((\partial h)^2 e^h)\}&=&
   \{\Q,\,C_{\rm L}((\partial h)^2(e^{2h}-e^h))\},
\end{eqnarray}
and others are zero.
We define similar operators $Q_{\rm R}(f)$ and $C_{\rm R}(f)$ by
replacing the contour $C_{\rm left}$ in (\ref{Eq:QCdef}) to $C_{\rm
right}$ corresponding to the right half of strings. Then, for the star
product, these operators satisfy
\begin{eqnarray}
\label{Eq:QC2}
 \left(Q_{\rm R}(e^h-1)\,A\right)*B=-(-1)^{\abs{A}}\,
A*\left(Q_{\rm L}(e^h-1)\,B\right), \nn
 \left(C_{\rm R}((\partial h)^2 e^h)\,
A\right)*B=-(-1)^{\abs{A}}\,A*\left(
C_{\rm L}((\partial h)^2 e^h)\,B\right).
\end{eqnarray}
Through conservation of the BRS current and the ghost field on the
identity string field, we find that
\begin{eqnarray}
\label{Eq:QC3}
  (Q_{\rm L}(e^h-1)+
Q_{\rm R}(e^h-1))\,I=0,\ \ \ 
  (C_{\rm L}((\partial h)^2 e^h)+
C_{\rm R}((\partial h)^2 e^h))\,I=0.
\end{eqnarray}
From (\ref{Eq:QC1}), (\ref{Eq:QC2}) and (\ref{Eq:QC3}),
it follows that
\begin{eqnarray}
 \Q\Psi_0&=&-\{\Q,\,C_{\rm L}((\partial h)^2 e^h)\}I, \nn
 \Psi_0*\Psi_0&=&
  \left(Q_{\rm L}(e^h-1)-C_{\rm L}((\partial h)^2e^h)\right)^2 I
   = \{\Q,\,C_{\rm L}((\partial h)^2e^h)\} I.
\end{eqnarray}
As a result, the equation of (\ref{Eq:solution}) is a classical solution
in the string field theory.

Though
the function $h(w)$ must cancel the midpoint singularity of
the ghost field on $I$
to make the operator $C_{\rm L}\,I$ well-defined,
this cancellation actually occurs
due to the previous two
conditions for $h(w)$.
Around the midpoint $w_0=\pm i$,
the ghost field behaves as \cite{rf:TT,rf:TT2,rf:Schnabl2}
\begin{eqnarray}
 c(w)I \sim \frac{1}{w-w_0}\left(-c_0+\frac{w_0}{2}(c_1-c_{-1})
\right)I+{\rm O}((w-w_0)^0),
\end{eqnarray}
and then its singularity is a pole at the midpoint.
If the function $h(w)$ is analytic around $w_0$, $h(w)$ can be expanded
as $h(w)=h''(w_0)(w-w_0)^2+\cdots$ because
$h'(w)=h'(-1/w)/w^2$. Therefore, $(\partial
h)^2 e^h\,c(w)$ becomes 
regular at the midpoint and then the operator $C_{\rm L}$ is
well-defined on the identity string field.

In the following, let us consider the solution generated by the function
\begin{eqnarray}
\label{Eq:ha}
h_a(w)=\log\left(1+\frac{a}{2}\left(w+\frac{1}{w}\right)^2\right),
\end{eqnarray}
and we parameterize the solution by a real number $a$.
The function is rewritten as $h_a(\sigma)=\log(1+2a\cos^2\sigma)$
on the unit circle $w=\exp(i\sigma)$, and
the parameter $a$ is
larger than $-1/2$ accordingly.
In the region $a\geq-1/2$, the function $h_a(\sigma)$
can be expanded by the Fourier series
\begin{eqnarray}
\label{Eq:haexp}
 h_a(\sigma)=-\log(1-Z(a))^2-2
\sum_{n=1}^\infty\frac{(-1)^n}{n}Z(a)^n \cos(2n\sigma),
\end{eqnarray}
where $Z(a)=(1+a-\sqrt{1+2a})/a$ and $-1\leq Z(a)<1\ (-1/2\leq a<\infty)$.

\begin{figure}[h]
 \epsfxsize=9.5cm
 \centerline{\epsfbox{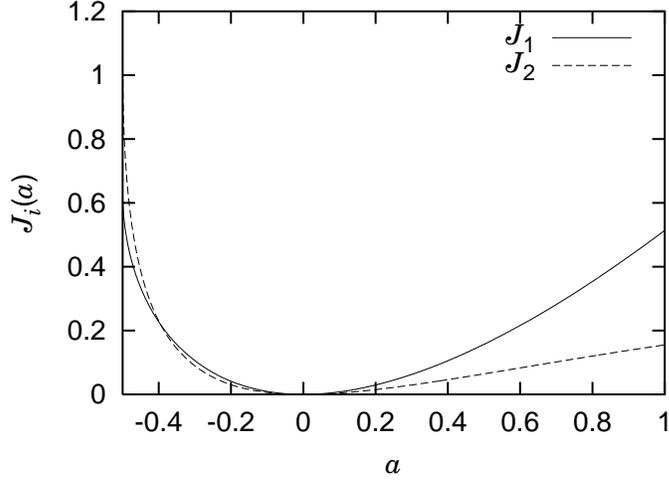}}
 \caption{Plots of the functions $J_1(a)$ and $J_2(a)$.} 
\label{fig:J}
\end{figure}

Substituting the function into the form (\ref{Eq:solution}) and
expanding it by oscillators, we find the solution up to level two
\begin{eqnarray}
\label{Eq:expsol}
 \ket{\Psi_0(a)}&=&J_1(a)\,c_1\ket{0}
+\left(\frac{8a}{3\pi}+J_1(a)\right)L_{-2}^X\,c_1\ket{0}\nn
&&+\left(\frac{8a}{\pi}+J_2(a)\right)c_{-1}\ket{0}
+\left(\frac{8a}{3\pi}+2J_1(a)\right)c_0\,b_{-2}\,c_1\ket{0}
+\cdots,
\end{eqnarray}
where $L_n^X$ denote matter Virasoro generators and 
$J_1(a)$ and $J_2(a)$ are given by
\begin{eqnarray}
 J_1(a) &=& -\int_{-\frac{\pi}{2}}^{\frac{\pi}{2}}
\frac{d\sigma}{2\pi}\,h_a'(\sigma)^2\,e^{h_a(\sigma)}\,
\frac{1}{2\cos\sigma}, \nn
J_2(a) &=& -\int_{-\frac{\pi}{2}}^{\frac{\pi}{2}}
\frac{d\sigma}{2\pi}\,
\,h_a'(\sigma)^2\,e^{h_a(\sigma)}\,
\frac{1+2\cos(2\sigma)}{2\cos\sigma}.
\end{eqnarray}
Using the Fourier series (\ref{Eq:haexp}), we can carry out these
integrations.
The results of the calculations are, for $a\geq 0$
\begin{eqnarray}
\hspace{-1cm}
&&
J_1(a)=
-\frac{4a}{\pi}\left\{
 1-\frac{1}{2}\left(\sqrt{Z(a)}+\frac{1}{\sqrt{Z(a)}}\right)
 \log \frac{1+\sqrt{Z(a)}}{1-\sqrt{Z(a)}}
\right\},\\
\hspace{-1cm}
&&
J_2(a) =
\frac{4a}{\pi}
\left\{
\frac{1}{3}+Z(a)+\frac{1}{Z(a)}
-\frac{1}{2}\left(Z(a)\sqrt{Z(a)}+\frac{1}{Z(a)\sqrt{Z(a)}}\right)
 \log \frac{1+\sqrt{Z(a)}}{1-\sqrt{Z(a)}}
\right\},
\end{eqnarray}
and for $-1/2\leq a <0$,
\begin{eqnarray}
\hspace{-1.3cm}
&&
J_1(a)=
-\frac{4a}{\pi}\left\{
 1+\left(\sqrt{-Z(a)}-\frac{1}{\sqrt{-Z(a)}}\right)
 \arctan \sqrt{-Z(a)}
\right\},\\
\hspace{-1.3cm}
&&
J_2(a) =
\frac{4a}{\pi}
\left\{
\frac{1}{3}+Z(a)+\frac{1}{Z(a)}
+\left(Z(a)\sqrt{-Z(a)}-\frac{1}{Z(a)\sqrt{-Z(a)}}\right)\arctan\sqrt{-Z(a)}
\right\}.
\end{eqnarray}
It is interesting to note that
the solution has a well-defined
Fock space expression and  
the coefficients of its component states have no divergence. This
situation is different from the 
case of the dilaton condensation in light-cone type string field
theories \cite{rf:Yoneya,rf:HN}. For instance, the functions
$J_1(a)$ and $J_2(a)$ have finite values as depicted in
Fig.~\ref{fig:J} and then the coefficients become finite up to level two. 
Moreover, the Fock space used in the solution can be spanned by the
universal basis, because the
solution is made of the BRS current, the ghost field and the identity
string field. This universality is indispensable
for the tachyon solution \cite{rf:SenUniv}. 
Finally, we indicate that
the solution is outside Siegel gauge since it contains
states proportional to the ghost zero mode $c_0$.

\subsection{physical interpretation}

To consider physical meaning of the solution, we expand
the string field $\Psi$ by the solution $\Psi_0$ and the quantum
fluctuation $\Phi$ as
\begin{eqnarray}
 \Psi=\Psi_0(a)+\Phi.
\end{eqnarray}
Substituting this form into (\ref{Eq:action}), 
the action becomes
\begin{eqnarray}
\label{Eq:shift-action}
 S[\Psi]=S[\Psi_0(a)]-\frac{1}{g^2}\int
\left(\frac{1}{2}\Phi*\Q'(a)\Phi+
\frac{1}{3}\Phi*\Phi*\Phi\right),
\end{eqnarray}
where the modified BRS charge $\Q'$ is given by
\begin{eqnarray}
\label{Eq:newBRS}
 \Q'(a)=Q(e^{h_a})-C((\partial h_a)^2 e^{h_a}),
\end{eqnarray}
and $Q(f)$ and $C(g)$ are defined by $Q_{\rm L}(f)+Q_{\rm R}(f)$
and $C_{\rm L}(g)+C_{\rm R}(g)$, respectively. The first term
in the action corresponds to the vacuum energy at the solution,
and the second term represents
the action for the quantum fluctuation.

We can show that the solution for $a>-1/2$ is expressed by a
gauge transformation of the trivial vacuum as
\begin{eqnarray}
\label{Eq:solgauge}
 \Psi_0(a)=\exp(K_{\rm L}(h_a)I)*\Q \exp(-K_{\rm L}(h_a)I),
\end{eqnarray}
where $\exp A$ for a string field $A$ is defined by its series as
$\exp A=I+A+A*A/2!+\cdots$, and the operator $K_{\rm L}$ is defined by
\begin{eqnarray}
 K_{\rm L}(f)=\int_{C_{\rm left}}
\frac{dw}{2\pi i}\,f(w) \left(J_{\rm gh}(w) 
 -\frac{3}{2}w^{-1}\right).
\end{eqnarray}
Consequently, we naturally expect that the theory around the solution
for $a>-1/2$ describes the physics on the perturbative
vacuum. Indeed, we can transform the action for the fluctuation $\Phi$
into the original action through the string field redefinition
\begin{eqnarray}
\label{Eq:redef}
 \Phi'=e^{K(h_a)} \Phi,
\end{eqnarray}
where $K(f)=K_{\rm L}(f)+K_{\rm R}(f)$, and $K_{\rm R}(f)$ is the counterpart
of $K_{\rm L}(f)$ related to right strings. The equivalence of these
actions is based on the fact that the original and
modified BRS charges are connected by the similarity transformation
\begin{eqnarray}
\Q'(a)=e^{K(h_a)}\Q e^{-K(h_a)}. 
\end{eqnarray}

However, since the operators $e^{K_{\rm L}}$ and $e^K$ becomes
ill-defined at $a=-1/2$,
the solution can not be represented by the pure
gauge, and
the theory around it can not be connected to the
original theory through the string field redefinition.
Consequently, 
the solution  at $a=-1/2$ represents a non-trivial 
vacuum, while the solution is a pure gauge for $a>-1/2$.
For example, the operator $e^K$ is written by the normal ordered form
\begin{eqnarray}
\label{Eq:expKnorm}
 e^{K(h_a)}&=&(1-Z(a)^2)^{-1}\exp(-\tilde{q}_0\log(1-Z(a))^2)\nn
 &&\times \exp\left(-\sum_{n=1}^\infty\frac{(-1)^n}{n}q_{-2n}Z(a)^n\right)
      \exp\left(-\sum_{n=1}^\infty\frac{(-1)^n}{n}q_{2n}Z(a)^n\right),
\end{eqnarray}
where $q_n$ denote the oscillators of the ghost number current and they
are written by the ghost oscillators as
\begin{eqnarray}
 && \tilde{q}_0=\frac{1}{2}(c_0 b_0-b_0 c_0) +\sum_{n=1}^\infty
   \left(c_{-n}b_n-b_{-n}c_n\right), \nn
 && q_n=\sum_{m=-\infty}^\infty c_{n-m}b_m\ \ \ (n\neq 0).
\end{eqnarray}
If we take $a=-1/2$, the first factor in (\ref{Eq:expKnorm}) diverges
because $Z(-1/2)=-1$. 
In order to find the singularity more explicitly, we write the string
field by the oscillator expression as
\begin{eqnarray}
 \ket{\Psi}=\phi(x)\,c_1 \ket{0}+\cdots
+\beta(x)\,c_{-1}\ket{0}+\gamma(x)\,b_{-2}\,c_0\,c_1
\ket{0}+\cdots.
\end{eqnarray}
Using the normal ordered expression (\ref{Eq:expKnorm}),
we find that, through the redefinition of (\ref{Eq:redef}), the
lowest level component field $\phi(x)$ is transformed as
\begin{eqnarray}
 \phi'(x)=\frac{1}{1+Z(a)}\phi(x)+\frac{Z(a)}{1+Z(a)}(-\beta(x)+2\gamma(x))
+\cdots,
\end{eqnarray}
where the abbreviation denotes the contribution from the higher level
component fields. Thus, by the string field redefinition, the component
fields are transformed into the linear combination of an infinite
number of fields. However, its coefficients
diverge at $a=-1/2$ and then this redefinition is
ill-defined. Similarly, the
operator $e^{K_{\rm L}}$ has a singularity at $a=-1/2$.

To find the physical meaning of the solution at $a=-1/2$ from a
different viewpoint,
we can determine the cohomology 
of the new BRS charge 
and the perturbative spectrum around
the solution. As in ref.~\cite{rf:KT}, the new cohomology
appears only in the sector with different ghost numbers from the original
cohomology. Consequently, in the gauge unfixed theory around the
solution, we can solve the equation of motion $\Q'(-1/2)\Phi=0$ as
$\Phi=\Q'(-1/2)\phi$.
Since the theory is invariant under the gauge transformation $\delta
\Phi=\Q'(-1/2) \delta\Lambda$, 
all of the on-shell physical states become gauge degrees of freedom in the
theory perturbatively.

Hence, we find that the universal solution at $a=-1/2$ represents
a non-trivial vacuum, where there is no physical excitation
perturbatively. This is the very feature required for
the tachyon vacuum. As discussed above, the solution satisfies universality.
Putting these observations together, we expect that the
universal solution corresponds to the tachyon vacuum itself.

\subsection{potential heights}

Formally, we can find that the potential height $-S[\Psi_0(a)]$ is zero for
$a>-1/2$.\footnote{More precisely, the potential $V$ is
divided by the D-brane volume $V_{\rm D}$
as $V=-S[\Psi_0]/V_{\rm D}$.}
Indeed, the derivative of $S[\Psi(a)]$ with respect to $a$
is given by
\begin{eqnarray}
 \frac{d}{da}S[\Psi_0(a)]=-\frac{1}{g^2}\int
 (\Q\Psi_0(a)+\Psi_0(a)*\Psi_0(a))*\frac{d\Psi_0(a)}{da}=0,
\end{eqnarray}
where we have used the equation of motions for the last
equality \cite{rf:KZ}. Since $h_{a=0}=\partial h_{a=0}=0$ and then
$S[\Psi_0(a=0)]=0$, it turns out that $S[\Psi_0(a)]$ is equal to zero. 
This zero potential height
can be shown only for $a>-1/2$, because the solution is ill-defined for
$a<-1/2$ 
and it is undifferentiable at $a=-1/2$.
This formal discussion is consistent with the expectation that the
universal solution is a pure gauge for $a>-1/2$.

However, we can not calculate the potential height more explicitly
beyond the formal evaluation.
Substituting the solution (\ref{Eq:solution}) into the action
(\ref{Eq:action}), we find that
\begin{eqnarray}
 S[\Psi_0(a)]&=&-\frac{1}{6g^2}\bra{I}C_{\rm L}((\partial h_a)^2 e^{h_a})\,
\Q\,C_{\rm L}((\partial h_a)^2 e^{h_a})\ket{I}\nn
&=&-\frac{1}{6g^2}
\int_{C_{\rm left}}\frac{dw}{2\pi i}
\int_{C_{\rm left}}\frac{dw'}{2\pi i}\nn
&&\times
(\partial h_a(w))^2 e^{h_a(w)}\,(\partial h_a(w'))^2 e^{h_a(w')}
\bra{I}c(w)\, \partial c(w')\ket{I},
\end{eqnarray}
where we have used
$\Q I=0$ and $\{\Q,\,c\}=c\partial c$. The identity string field is
written by the tensor product of the matter and ghost sectors and the
matter sector of the identity string field is given by
\cite{rf:GJ}
\begin{eqnarray}
 \ket{I^X}=\exp\left(
-\sum_{n=1}^\infty\frac{(-1)^n}{2n}\alpha_{-n}\cdot
\alpha_{-n}\right)\ket{0}.
\end{eqnarray}
Then, $\bra{I}c\partial c\ket{I}$ is an indefinite quantity
for any $a$
due to the infinite determinant factor of the matter sector.
Thus, it is difficult to evaluate the potential height
because we have not yet known how it should be regularized\footnote{
We are still suffering from the disastrous divergence even if we use
usual regularization with the insertion of $e^{-\epsilon L_0}$
\cite{rf:KTunpub}.}. 

Instead of the exact calculation, we try to use level truncated
solutions to evaluate the potential height.
For $a=-1/2$,
the solution (\ref{Eq:expsol}) becomes
\begin{eqnarray}
\label{Eq:expsol2}
 \ket{\Psi_0(-1/2)}&=&\frac{2}{\pi}\,c_1\ket{0}
+\frac{2}{3\pi}L_{-2}^X\,c_1\ket{0}
-\frac{2}{3\pi}\,c_{-1}\ket{0}
+\frac{8}{3\pi}\,c_0\,b_{-2}\,c_1\ket{0}
+\cdots.
\end{eqnarray}
At level zero, the truncated solution is $2/\pi\times c_1\ket{0}$ and so
the function $f(T)$ defined in 
ref.~\cite{rf:SZ-tachyon} 
takes the value
\begin{eqnarray}
 2\pi^2\left(
-\frac{1}{2}\left(\frac{2}{\pi}\right)^2
+\frac{1}{3}\left(\frac{2}{\pi}\right)^3\right)
\simeq -0.279.
\end{eqnarray}
This provides 28\% of the D-brane tension.
Furthermore, we calculate the function $f(T)$ at level two and, then,
it takes the value $\sim 85$, which is far from a stationary point at
the level two potential.
For general $a$, we are faced with such terrible behavior.
Although this result discourages us to perform further
calculations,
this is a natural result because our solution is not a
solution in the level truncated theory.

Hence, we can not calculate the vacuum energy of the universal
solutions at present, in order to compare it with the D-brane tension.

\section{Level truncation in string field theory with the modified
kinetic term} 

In this section we explore another possibility of clarifying the relation
between the universal solutions and the tachyon vacuum.
Instead of the direct calculation of the potential height,
we analyze the non-perturbative vacuum
in the theory expanded around the universal solutions.
By using the level truncation scheme in Siegel gauge, 
we can find the non-perturbative vacuum without any difficulty,
and moreover we come across the impressive result which supports our
conjecture for the universal solutions.

\subsection{conjectures and setup}

As in (\ref{Eq:shift-action}), the action for the fluctuation $\Phi$
is written by
\begin{eqnarray}
\label{Eq:action-fluc}
 S[\Phi]=-\frac{1}{g^2}\int
\left(\frac{1}{2}\Phi*\Q'(a)\Phi
+\frac{1}{3}\Phi*\Phi*\Phi\right),
\end{eqnarray}
where the modified BRS charge is given by (\ref{Eq:newBRS}).
Substituting
(\ref{Eq:haexp}) into (\ref{Eq:newBRS}) and performing the
$w$ integration, 
we obtain the oscillator expressions of the new BRS charge
\begin{eqnarray}
\label{Eq:newBRS2}
 \Q'(a)&=& (1+a)\Q+\frac{a}{2}(Q_2+Q_{-2})
+4a Z(a)\,c_0-2a Z(a)^2 (c_2+c_{-2})\nn
&&-2a (1-Z(a)^2) \sum_{n=2}^\infty (-1)^n Z(a)^{n-1}(c_{2n}+c_{-2n}),
\end{eqnarray}
where we expand the BRS current as $J_{\rm B}(w)=\sum_n
Q_n w^{-n-1}$.
The details of the calculation are presented in Appendix A. 

Under the Siegel gauge condition $b_0 \Phi=0$,
the quadratic term in the action becomes
\begin{eqnarray}
&&-\frac{1}{g^2} \int \frac{1}{2}\,\Phi*\Q'(a)\,\Phi
= -\frac{1}{g^2}\int \frac{1}{2}\,\Phi*c_0\,L(a)\,\Phi,
\end{eqnarray}
where $L(a)= \left\{\Q'(a),\,b_0\right\}$.
Using the anti-commutation relation \cite{rf:KT}
\begin{eqnarray}
 \left\{Q_m,\,b_n\right\}=L_{m+n}+mq_{m+n}
+\frac{3}{2}m(m-1)\delta_{m+n,0},
\end{eqnarray}
we can calculate the operator $L(a)$ as
\begin{eqnarray}
\label{Eq:La}
L(a)= (1+a)L_0 + \frac{a}{2}(L_2+L_{-2})+a\,(q_2-q_{-2})+4a Z(a). 
\end{eqnarray}
Therefore, according to the notations of
\cite{rf:SenUniv,rf:SZ-tachyon}, the potential in the new string field
theory is given 
by the `modified' universal function
\begin{eqnarray}
 f_a(\Phi)=2\pi^2\left(\frac{1}{2}
\left<\Phi,c_0\,L(a)\Phi\right>
+\frac{1}{3}\left<\Phi,\Phi*\Phi\right>\right).
\end{eqnarray}
As seen in the previous section, we expect that 
the solutions $\Psi_0(a>-1/2)$ and $\Psi_0(a=-1/2)$ 
are regarded as a pure gauge and the tachyon vacuum,
respectively. Consequently, in the case of $a>-1/2$, the action of
(\ref{Eq:action-fluc}) describes the perturbative vacuum,
and then, in the potential, there is the stationary point which
corresponds to the tachyon vacuum.
On the other hand, 
the stationary point must vanish at $a=-1/2$,
because 
the theory has already stayed on the non-perturbative
vacuum. Hence, due to our conjectures, 
the modified universal function
at the stationary point $\Phi_0$ must satisfy
\begin{eqnarray}
\label{Eq:fastep} f_a(\Phi_0)=\left\{
\begin{array}{rl}
 0& (a=-1/2)\\
 -1& (a>-1/2).
\end{array}\right.
\end{eqnarray}

Let us consider a level truncated expression of the modified universal
function. 
The new action is invariant under the twist
transformation $\sigma\rightarrow
\pi-\sigma$ \cite{rf:SZ-tachyon}.  Due to this symmetry, we have only
to look for a stationary point where $\Phi_0$ contains even level states
as well as the case of the original level truncated analysis.
In general, we can write a scalar string field $\ket{\Phi}$ by the tensor
product of the matter and ghost states as
\begin{eqnarray}
\ket{\Phi}=\sum_{i=0}^\infty \psi_i \ket{s_i},\ \ \ 
\ket{s_i}=|\eta_{m(i)}\rangle\otimes |\chi_{g(i)}\rangle.
\end{eqnarray}
Our notations for the decomposition of states are almost same as
in Ref.~\cite{rf:MT}. Up to level 6, the matter and ghost states,
$\ket{\eta_i}$ and $\ket{\chi_i}$, are given
in Appendix B and the decomposition of states is in Appendix C.

Using the component fields $\psi_i$ with zero momentum,
we can express the modified universal function as follows,
\begin{eqnarray}
\label{Eq:eqpsi}
 f_a(\psi)=2\pi^2
\left(\frac{1}{2}\sum_{ij}d_{ij}(a)\psi_i\psi_j
+\frac{1}{3}\sum_{ijk}t_{ijk}\psi_i\psi_j\psi_k\right).
\end{eqnarray}
The cubic coefficients $t_{ijk}$ does not change from the previous
analysis in Ref.~\cite{rf:MT}.
The quadratic coefficients are calculated as
\begin{eqnarray}
\label{Eq:dij}
d_{ij}(a)&=&\left\{(1+a)({\rm level}(i)-1)
A_{ij}^{\rm mat}A_{ij}^{\rm gh}
+4aZ(a)\right\}A_{ij}^{\rm mat}A_{ij}^{\rm gh} \nn
&&
+a B_{ij}^{\rm mat}A_{ij}^{\rm gh}+a A_{ij}^{\rm mat}B_{ij}^{\rm gh},
\end{eqnarray}
where ${\rm level}(i)$ denotes 
the level of the state $\ket{s_i}$,\footnote{We set the level of ground
states as ${\rm
level}(\eta_0)=0$ and ${\rm level}(\chi_0)=0$.}
and $A_{ij}^{\rm mat(gh)}$ and $B_{ij}^{\rm mat(gh)}$ are defined by
\begin{eqnarray}
\label{Eq:Amat}
&& A_{ij}^{\rm mat}=\langle\eta_{m(i)} | \eta_{m(j)}\rangle,\\
\label{Eq:Agh}
&& A_{ij}^{\rm gh}=\langle\chi_{g(i)} | c_0 |\chi_{g(j)}\rangle,\\
\label{Eq:Bmat}
&& B_{ij}^{\rm mat}=\langle\eta_{m(i)} | L_{-2}^{\rm mat}|
\eta_{m(j)}\rangle,\\
\label{Eq:Bgh}
&& B_{ij}^{\rm gh}=\langle\chi_{g(i)} | c_0\, 
(L_{-2}^{\rm gh}-q_{-2})|
\chi_{g(j)}\rangle.
\end{eqnarray}
Here, $L_n^{\rm mat}$ and $L_n^{\rm gh}$ are the matter and ghost parts
of the total Virasoro generators $L_n$. In this expression,
 $L_2$ and $q_2$ are converted to $L_{-2}$ and $-q_{-2}$
by the hermitian conjugation.
A list of the coefficients $A_{ij}^{\rm mat(gh)}$ up to level 6 is given
in Appendix B. Up to level 6, the coefficients $B_{ij}^{\rm
mat(gh)}$ can be calculated by using $A_{ij}^{\rm mat(gh)}$ through the
following equations, 
\begin{eqnarray}
\label{Eq:L2state}
 L^{\rm mat}_{-2} \ket{\eta_0} &=& \half \ket{\eta_1}, \nn
 L^{\rm mat}_{-2} \ket{\eta_1} &=& 2 \ket{\eta_3} + \half \ket{\eta_5},\nn
 L^{\rm mat}_{-2} \ket{\eta_3} &=& 3 \ket{\eta_6} + \ket{\eta_8}
                             +\half \ket{\eta_9},\nn
 L^{\rm mat}_{-2} \ket{\eta_4} &=& 4 \ket{\eta_7} + \ket{\eta_{10}}, \nn
 L^{\rm mat}_{-2} \ket{\eta_5} &=& 4 \ket{\eta_9} + \half \ket{\eta_{12}}, \nn
 \left(L^{\rm gh}_{-2}-2 q_{-2}\right) \ket{\chi_0} &=&
   -\ket{\chi_1}, \nn
 \left(L^{\rm gh}_{-2}-2 q_{-2}\right) \ket{\chi_1} &=&
   3\ket{\chi_4} +\ket{\chi_6}, \nn
 \left(L^{\rm gh}_{-2}-2 q_{-2}\right) \ket{\chi_4} &=&
  5\ket{\chi_7} + \ket{\chi_9}, \nn
 \left(L^{\rm gh}_{-2}-2 q_{-2}\right) \ket{\chi_5} &=&
  4\ket{\chi_8}+2\ket{\chi_{10}}+\ket{\chi_{12}}, \nn
 \left(L^{\rm gh}_{-2}-2 q_{-2}\right) \ket{\chi_6} &=&
  3\ket{\chi_9}+3\ket{\chi_{11}}.
\end{eqnarray}

\subsection{level zero analysis}

\begin{figure}[ht]
 \epsfxsize=9cm
 \centerline{\epsfbox{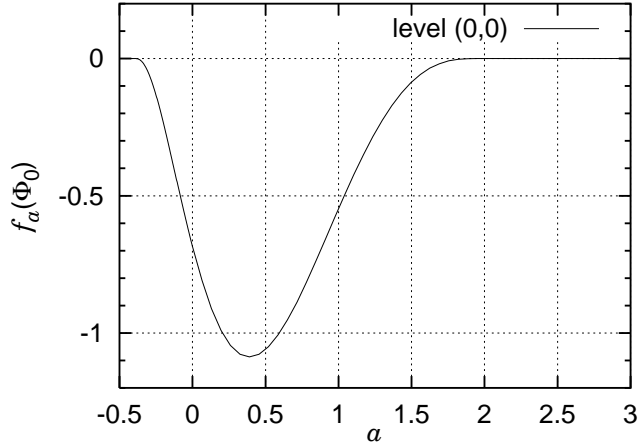}}
 \caption{Plots of the potential minimum in level
 $(0,0)$ truncation.} 
\label{fig:level0}
\end{figure}

At level $(0,\,0)$ approximation,
the component field is $t\,c_1\ket{0}$ and then the modified universal
function is 
\begin{eqnarray}
 f_a(t) = 2\pi^2\left(-\frac{1}{2}\lambda(a)\,t^2
+\frac{1}{3}K^3\,t^3 \right),
\end{eqnarray}
where $\lambda(a)=4\sqrt{1+2a}-3(1+a)$ and $K=3\sqrt{3}/4$.
It is easy to see that $\lambda(a)$ has two roots
\begin{eqnarray}
 a^+=\frac{7+4\sqrt{7}}{9}=1.954\cdots,\ \ \ 
 a^-=\frac{7-4\sqrt{7}}{9}=-0.398\cdots,
\end{eqnarray}
then
\begin{eqnarray}
&& {\rm if}\ \ a^-<a<a^+,\hspace{3.2cm} \lambda(a)>0,\nn
&& {\rm if}\ \ \ -1/2\leq a<a^-\ {\rm or}\ a>a^+,\hspace{.5cm} \lambda(a)<0.
\end{eqnarray}
Therefore, $f_a(t)$ has a local minimum at
\renewcommand{\arraystretch}{1.5}
\begin{eqnarray}
t_0=
\left\{
\begin{array}{cl}
\displaystyle
K^{-3}\,\lambda(a) &(a^-\leq a\leq a^+)\\
\displaystyle
0 &(-1/2\leq a<a^-\ {\rm or}\ a>a^+),
\end{array}
\right.
\end{eqnarray}
\renewcommand{\arraystretch}{1}
and $f_a(t)$ at this minimum takes the value
\renewcommand{\arraystretch}{2}
\begin{eqnarray}
f_a(t_0)=
\left\{
\begin{array}{cl}
\displaystyle
-\frac{\pi^2}{3 K^6}\,\lambda(a)^3 &(a^-\leq a\leq a^+)\\
\displaystyle
0 &(-1/2\leq a<a^-\ {\rm or}\ a>a^+).
\end{array}
\right.
\end{eqnarray}
\renewcommand{\arraystretch}{1}
The $a$ dependence of this value is depicted in
Fig.~\ref{fig:level0}. Though this behavior is quite different from the
expectation that $a$ does not affect the potential minimum for $a>-1/2$
as in (\ref{Eq:fastep}), this $a$ dependence is introduced merely by the
level truncation approximation.

The value $-0.684\cdots$ at $a=0$ equals to the minimum
derived from the previous level truncation analysis \cite{rf:SZ-tachyon},
because the kinetic 
operator $L(a)$ becomes $L_0$ at $a=0$ and the modified
universal function agrees with the universal function. This agreement is
realized for any level analysis as seen in (\ref{Eq:La}). In addition,
it should be noticed that the minimum is exactly zero at $a=-1/2$.

\subsection{higher level analysis}

We apply the iterative approximation algorithm used by Moeller and
Taylor \cite{rf:MT} to higher level calculations.
However, there is a slight change
in the procedure to find the stable vacuum. 
According to Ref.~\cite{rf:MT},
the solution which minimizes the potential of (\ref{Eq:eqpsi})
has many branches, and which branch should be chosen is determined by the
condition that the solution
becomes the level zero stable vacuum if higher level fields are turned
off. As a result, the branch depends on 
the sign of the quadratic coefficients $d_{ij}(a)$.
As an example, let us see the tachyon field $\psi_0=t$.
Since the coefficient $d_{00}(a)$ equals to
$-\lambda(a)$ for any level analysis, 
the tachyon field which minimizes the potential
can be expressed by other
fields in the following,
\begin{eqnarray}
 t=
\left\{
\renewcommand{\arraystretch}{2.2}
\begin{array}{ll}
\displaystyle
\frac{-\beta+\sqrt{\beta^2-4\alpha\gamma}}{2\alpha} 
& (a^-\leq a\leq a^+)\\
\displaystyle
\frac{-\beta-\sqrt{\beta^2-4\alpha\gamma}}{2\alpha} 
& (-1/2\leq a<a^-,\ \ a>a^+).
\end{array}
\renewcommand{\arraystretch}{1}
\right.
\end{eqnarray}
Here, $\alpha$, $\beta$ and $\gamma$ are given by
\begin{eqnarray}
&& \alpha=t_{000}, \nn
&& \beta=-\lambda(a)+\sum_{i=1}^N t_{00i}\psi_i, \\
&& \gamma=\sum_{i=1}^Nd_{0i}(a)\psi_i
+\sum_{i,j=1}^N t_{0ij}\psi_i\psi_j,
\end{eqnarray}
where 
$N$ is the number of truncated fields.
Thus, the branch is determined depending on the
value of $a$ in our analysis.

\begin{figure}[h]
 \epsfxsize=10cm
 \centerline{\epsfbox{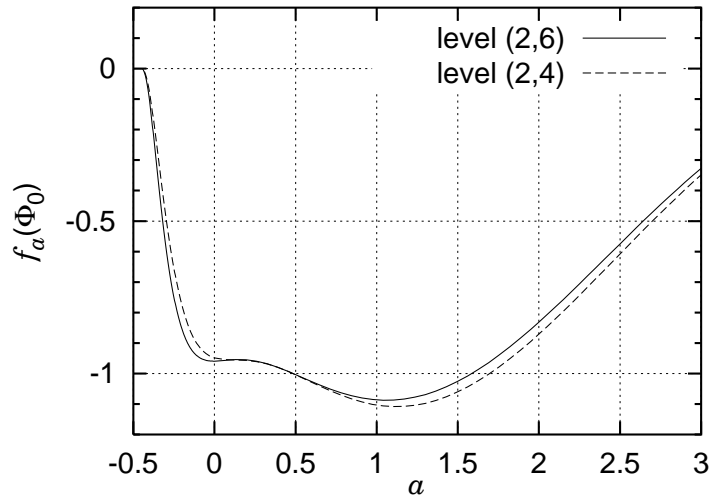}}
 \caption{The potential minimum in level two truncations.}
 \label{fig:level2}
\end{figure}

Let us consider the level two approximation. 
The level two field is given by
\begin{eqnarray}
 \ket{\Phi^{(2)}}&=&t \ket{s_0}+\psi_1 \ket{s_1}+\psi_2\ket{s_2}\\
&=& t\,c_1 \ket{0}+\psi_1\,(\alpha_{-1}\cdot\alpha_{-1})\,c_1\ket{0}
-\psi_2\,c_{-1}\ket{0}.
\end{eqnarray}
From (\ref{Eq:dij}) and (\ref{Eq:L2state}), we find the quadratic term of
the modified universal function as
\begin{eqnarray}
&& 2\pi^2\left(
  -\frac{1}{2}\lambda(a)\,t^2
+ 26\,(1 + a + 4 a Z(a))\,\psi_1^2-
\frac{1}{2}\,(1 + a + 4 a 
Z(a))\,\psi_2^2\right.\\
&&\left. + 
13 a\,t\,\psi_1 + \frac{1}{2}a\,t\,\psi_2 
\right).
\end{eqnarray}
The potential minimum is depicted in Fig.~\ref{fig:level2}.
We observe that $(2,4)$ and $(2,6)$ approximations lead to almost same
results. At $a=-1/2$, 
the vacuum expectation values of component fields
and the potential minimum are equal to zero
as used $(0,0)$ approximation. 
The vacuum energy are varying slowly in the neighborhood of $a=0$,
in which the potential height provide 96\% of the D-brane tension.
For our conjectures, it is a desirable fact that the potential minimum
changes slowly along the expected vacuum 
energy. Thus, even at this level, we can expect that the universal
solution is a pure gauge at least near $a=0$, as our conjecture.

\begin{figure}[t]
 \epsfxsize=9.3cm
 \centerline{\epsfbox{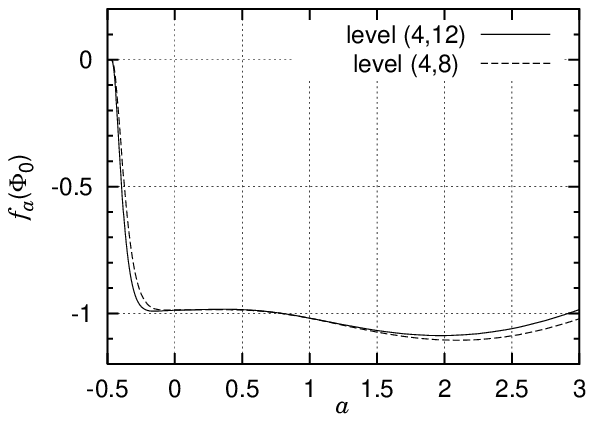}}
 \epsfxsize=9.3cm
 \centerline{\epsfbox{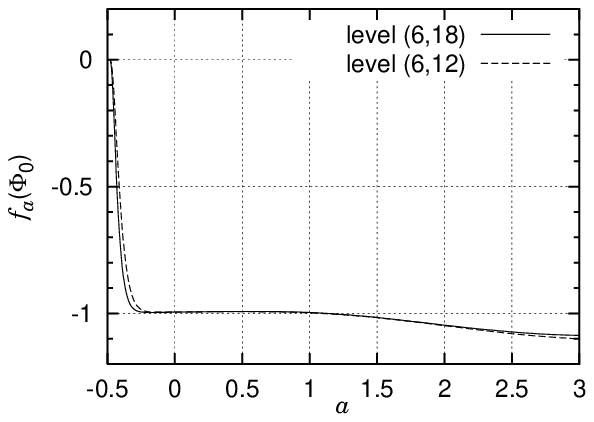}}
 \caption{Plots of the potential minimum in level
 four and six truncations.}
 \label{fig:level46}
\end{figure}

\begin{figure}[h]
\vspace{.5cm}
 \epsfxsize=9cm
 \centerline{\epsfbox{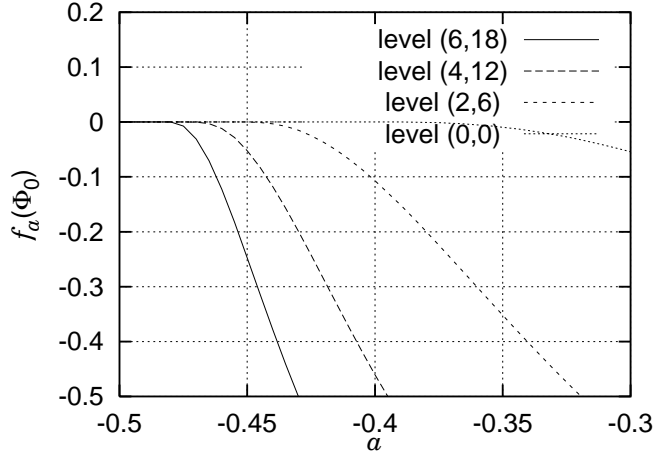}}
 \caption{The enlarged graph of the potential minimum around $a=-1/2$.}
 \label{fig:zoom}
\end{figure}

Let us progress to the higher level approximation. The
potential minimum using level four and six approximations is depicted in
Fig.~\ref{fig:level46}.
Like level zero and two cases,
the results of the level $(L,3L)$ calculation
slightly change with the level $(L,2L)$. In the cases of both of level
four and six, the potential minimum displays a flat region
along the expected vacuum energy.
The potential heights and the vacuum expectation values become zero at
$a=-1/2$ as before. This behavior can be seen 
in detail in Fig.~\ref{fig:zoom}, which magnifies the area near
$a=-1/2$.

In these analyses, it is a remarkable fact that
the higher the approximation level is increased, the wider the flat
region grows. Moreover, the potential value in the flat
region approaches the expected value increasingly
as the level is raised.
At level six, the vacuum energy becomes almost
$-1$ in the 
region from $-0.2$ to $1$.  In Fig.~\ref{fig:pot}, we pick out the
values of the potential 
height for several points of $a$.\footnote{Of course, our values at
$a=0$ agree with previous results in 
refs.~\cite{rf:SZ-tachyon,rf:MT,rf:GR}. But only the value at level
(6,12) does not coincide with a result in \cite{rf:MT}.}
All of the values are about 99\% of the D-brane tension at level
six. For our conjecture, the most important result is that the stable
vacuum disappears at $a=-1/2$ in every level analysis.
These potential behavior to the parameter $a$ suggests that, if
the approximation level approaches infinity,  the potential
minimum takes the value of $-1$ for $a>-1/2$, but
it remains being zero at $a=-1/2$.
Hence, these results lead us to believe 
that the conjecture for the universal solutions, which is expressed by
(\ref{Eq:fastep}), should be true, and then
the universal solution at $a=-1/2$
corresponds to the tachyon condensation conjectured by Sen.

\subsection{other universal solutions}

\begin{figure}[t]
\centerline{
\begin{tabular}{|l|r|r|r|r|r|}
\hline
 & \multicolumn{5}{c|}{$V/T_{25}$} \\
\cline{2-6}
\multicolumn{1}{|c|}{\raisebox{1.5ex}[0pt]{level}} &
\multicolumn{1}{c|}{\makebox[1.8cm]{$a=-0.2$}} &
\multicolumn{1}{c|}{\makebox[1.8cm]{$a=-0.1$}} &
\multicolumn{1}{c|}{\makebox[1.8cm]{$a=0.0$}} &
\multicolumn{1}{c|}{\makebox[1.8cm]{$a=0.1$}} &
\multicolumn{1}{c|}{\makebox[1.8cm]{$a=0.2$}} \\
\hline
\hline
(0,0) &
$-$0.233203 &
$-$0.462912 &
$-$0.684616 &
$-$0.866692 &
$-$0.995360 \\
\hline
(2,4) &
$-$0.777067 &
$-$0.908062 &
$-$0.948553 &
$-$0.955031 &
$-$0.956909 \\
\hline
(2,6) &
$-$0.854866 &
$-$0.944975 &
$-$0.959377 &
$-$0.955239 &
$-$0.955100 \\
\hline
(4,8) &
$-$0.965369 &
$-$0.986459 &
$-$0.986403 &
$-$0.985538 &
$-$0.985329 \\
\hline
(4,12) &
$-$0.988826 &
$-$0.990313 &
$-$0.987822 &
$-$0.986499 &
$-$0.985300 \\
\hline
(6,12) &
$-$0.993496 &
$-$0.995449 &
$-$0.994773 &
$-$0.994077 &
$-$0.993590 \\
\hline
(6,18) &
$-$0.996274 &
$-$0.996056 &
$-$0.995177 &
$-$0.994346 &
$-$0.993715 \\
\hline
\end{tabular}}
\caption{Vacuum energy in level truncation scheme for several points of $a$.}
\label{fig:pot}
\end{figure}

We can provide other universal solutions by choosing
the function in (\ref{Eq:solution}) as
\begin{eqnarray}
\label{Eq:solutionl}
h_a^l(w)=\log\left(
1-\frac{a}{2}(-1)^l\left(w^l-(-1)^l\,w^{-l}\right)^2\right),
\end{eqnarray}
where $l=1,\,2,\,3\,\cdots$  \cite{rf:KT}. The case of $l=1$ corresponds
to the previous example.
The action around the solution has the modified BRS charge
\begin{eqnarray}
\label{Eq:newBRS2l}
 \Q^l(a)&=& (1+a)\Q+(-1)^l\,\frac{a}{2}\,(Q_{2l}+Q_{-2l})
+4a l^2 Z(a)\,c_0+ (-1)^l\, a l^2\, Z(a)^2 (c_{2l}+c_{-2l})\nn
&&-2\,a l^2\, (1-Z(a)^2)
\sum_{n=2}^\infty (-1)^{nl} Z(a)^{n-1}(c_{2nl}+c_{-2nl}).
\end{eqnarray}
The kinetic operator in Siegel gauge is given by
\begin{eqnarray}
 L^l(a)&=&\{Q^l(a),\,b_0\}\\
&=& (1+a)L_0 -(-1)^l\,\frac{a}{2}\left(L_{2l}+L_{-2l}\right)
-(-1)^l\,al \,(q_{2l}-q_{-2l})+4al^2\,Z(a).
\end{eqnarray}
At level zero, the modified universal function becomes
\begin{eqnarray}
&&
 f_a^l(t)=2\pi^2\left(
-\frac{1}{2}\lambda_l(a)\,t^2+\frac{1}{3}\,K^3\,t^3\right),
\nn
&&
\lambda_l(a)=4l^2\sqrt{1+2a}-(4l^2-1)(1+a).
\end{eqnarray}
At the local minimum it takes the value
\renewcommand{\arraystretch}{2}
\begin{eqnarray}
f_a(t_0)=
\left\{
\begin{array}{cl}
\displaystyle
-\frac{\pi^2}{3 K^6}\,\lambda_l(a)^3 &(a_l^-\leq a\leq a_l^+)\\
\displaystyle
0 &(-1/2\leq a<a_l^-\ {\rm or}\ a>a_l^+),
\end{array}
\right.
\end{eqnarray}
\renewcommand{\arraystretch}{1}
where the branch points $a_l^\pm$ are given by
\begin{eqnarray}
 a_l^\pm = \frac{8l^2-1\pm 4l^2\sqrt{8l^1-1}}{(4l^2-1)^2}.
\end{eqnarray}

Let us consider the case of $l=2$. The kinetic operator is given by
\begin{eqnarray}
 L^2(a)=(1+a)L_0-\frac{a}{2}(L_4+L_{-4})
-2a(q_4+q_{-4})+16a\,Z(a).
\end{eqnarray}
Similarly to the case of $l=1$, the quadratic terms up to level six can be
calculated by the following equations,
\begin{eqnarray}
L_{-4}^{\rm mat} \ket{\eta_0}&=&
\ket{\eta_3}+\frac{1}{2}\ket{\eta_4},
\nn
L_{-4}^{\rm mat} \ket{\eta_1}&=&
\ket{\eta_9}+\ket{\eta_6}+\frac{1}{2}\ket{\eta_{10}},
\nn
\left(
L_{-4}^{\rm gh}-4q_{-4}\right)\ket{\chi_0}&=&
-3\ket{\chi_4}-2\ket{\chi_5}-\ket{\chi_6},
\nn
\left(
L_{-4}^{\rm gh}-4q_{-4}\right)\ket{\chi_1}&=&
5\ket{\chi_7}+\ket{\chi_{11}}+2\ket{\chi_{12}}.
\end{eqnarray}
The branch points are
\begin{eqnarray}
 a_2^-=-0.258\cdots,\ \ \ a_2^+=0.533\cdots.
\end{eqnarray}

\begin{figure}[h]
 \epsfxsize=9.5cm
 \centerline{\epsfbox{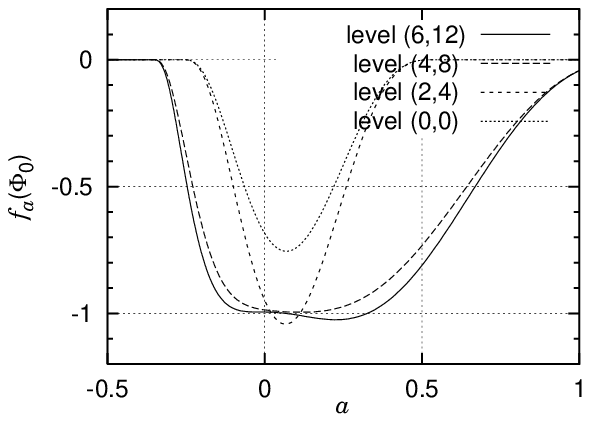}}
 \epsfxsize=9.5cm
 \centerline{\epsfbox{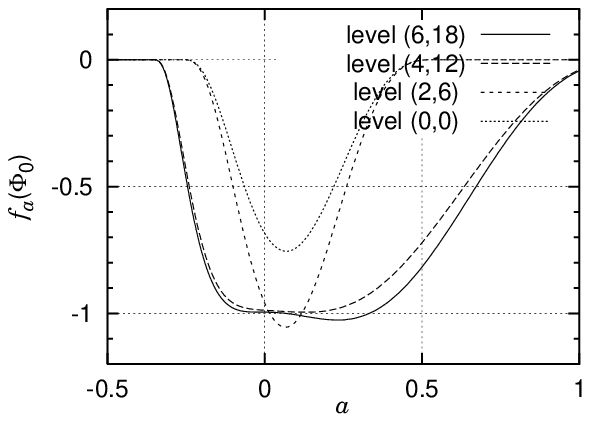}}
 \caption{The potential minimum for the universal solution with $l=2$.} 
 \label{fig:L2}
\end{figure}

The results of numerical analysis up to level six are depicted in
Fig.~\ref{fig:L2}.
At $a=-1/2$, the potential minimum remains being zero for any
approximation level. From level four, the flat region
begins to spread along the expected vacuum energy. As a different
feature from the previous case, we observe 
that, if the level is raised, the shape of the curve
changes with the period of level two,
in particular near the point that the
potential minimum approaches to zero.
On the whole we can see that the curve tends to approach the step
function (\ref{Eq:fastep}) as increasing the level.
However, it seems to converge more slowly than the case of $l=1$.

\section{Summary and discussions}

We have studied a non-perturbative vacuum in open string field theory 
around the universal solutions by using a level truncation scheme. We
have observed that, as increasing the approximation level,
the potential minimum gradually approaches to the step function which
equals to the negative D-brane tension for almost parameters and becomes
zero at the boundary. These results strongly support our conjecture that
the universal solution are 
pure gauge for the almost parameter region, but the non-trivial solution
at the boundary corresponds to the tachyon vacuum solution.

Though this physical interpretation for the universal solutions is
plausible, we can construct them as many as
functions used in eq.~(\ref{Eq:solution}).
At least, the non-trivial solutions are countable by the natural number
as in eq.~(\ref{Eq:solutionl}).
On the other hand  the tachyon vacuum should be unique in string theory.
In order for them to be consistent,
these universal solutions must be changed by gauge
transformations. Though this gauge equivalence
remains to be proved, this may be probably
possible by the gauge transformation related to the operators
$K_n=L_n-(-1)^n L_{-n}$ as discussed in \cite{rf:KT}. In addition, the
equivalence between our 
solutions in this paper is consistent with the conjecture proposed by
Drukker \cite{rf:Drukker1,rf:Drukker2} which concerns the order of zero
of the function with the number of D-branes.

Finally, two problems at least are remained to be solved
in order to prove the correspondence between the universal solutions and
the tachyon vacuum.
First,
we should calculate the potential height
of the universal solutions exactly instead of these numerical
analysis. Secondly, we should find closed 
strings in the theory around the solutions. In \cite{rf:Drukker2}
Drukker discussed how closed strings should be appeared in the
theory, but we should show them more explicitly, for example as closed
string poles in amplitudes.

\section*{Acknowledgements}
We would like to thank H.~Hata, M.~Kanemune, I.~Kishimoto and T.~Kugo
for valuable discussions and comments.

\newpage
\vspace{5ex}
\centerline{\Large\bf Appendix}
\appendix

\section{Mode expansion of the modified BRS charge}

The modified BRS charge $\Q'(a)$ is defined in (\ref{Eq:newBRS}) and the
function $h_a(w)$ is given by (\ref{Eq:ha}).
We can easily evaluate the first term of (\ref{Eq:newBRS}) as
\begin{eqnarray}
\label{Eq:Qfirst}
Q(e^{h_a}) &=& Q\left(1+\frac{a}{2}\left(w+\frac{1}{w}\right)^2\right)\nn
    &=& Q\left(1+a+\frac{a}{2}(w^2+w^{-2})\right)\nn
    &=& (1+a)\,\Q+\frac{a}{2}\,(Q_2+Q_{-2}).
\end{eqnarray}

From (\ref{Eq:ha}), we can find
\begin{eqnarray}
 (\partial h_a(w))^2 e^{h_a(w)} &=&
  a\,w^{-1}\,\partial h_a(w)\,(w^2-w^{-2}).
\end{eqnarray}
If we differentiate the Fourier series of (\ref{Eq:haexp}), we get
\begin{eqnarray}
 \partial h_a(w) &=& -4i\,w^{-1}\,
\sum_{n=1}^\infty (-1)^n Z(a)^n \sin 2n\sigma,
\end{eqnarray}
where $w=\exp(i\sigma)$. From these equations, it follows that
\begin{eqnarray}
\label{Eq:delh}
 (\partial h_a(w))^2 e^{h_a(w)} &=& 8\,a\,w^{-2}\,
\sum_{n=1}^\infty (-1)^n Z(a)^n \sin 2n\sigma\,\sin 2\sigma\nn
&=& 8\,a\,w^{-2}\,
\left(-Z(a)\sin^2 2\sigma+
\sum_{n=2}^\infty (-1)^n Z(a)^n \sin 2n\sigma\,\sin 2\sigma
\right).
\end{eqnarray}
Substituting (\ref{Eq:delh}) into (\ref{Eq:newBRS}), we can evaluate
the second term of (\ref{Eq:newBRS}) as 
\begin{eqnarray}
\label{Eq:Cha}
 C((\partial h_a)^2 e^{h_a}) &=& 
-8a Z(a) \sum_{n=-\infty}^\infty c_n
\int_{-\pi}^\pi \frac{d\sigma}{2\pi}\,e^{-in\sigma} \sin^2 2\sigma \nn
&&+8a \sum_{n=-\infty}^\infty c_n
\sum_{m=2}^\infty (-1)^m Z(a)^m
\int_{-\pi}^\pi \frac{d\sigma}{2\pi}\,e^{-in\sigma} \sin 2m\sigma
\sin 2\sigma.
\end{eqnarray}
We can calculate the integrations in this equation and the results are
\renewcommand{\arraystretch}{2}
\begin{eqnarray}
\label{Eq:int1}
\int_{-\pi}^\pi \frac{d\sigma}{2\pi}\,e^{-in\sigma}
\sin^2 2\sigma=
\left\{
\begin{array}{rl}
\displaystyle
\frac{1}{2} & (n=0)\\
\displaystyle
-\frac{1}{4} & (n=\pm 4)\\
\displaystyle
0 & ({\rm otherwise}),
\end{array}
\right.
\end{eqnarray}
\begin{eqnarray}
\label{Eq:int2}
\int_{-\pi}^\pi \frac{d\sigma}{2\pi}\,e^{-in\sigma}
\sin 2m\sigma \sin 2\sigma=
\left\{
\begin{array}{rl}
\displaystyle
\frac{1}{4} & (n=\pm 2(m-1))\\
\displaystyle
-\frac{1}{4} & (n=\pm 2(m+1))\\
\displaystyle
0 & ({\rm otherwise}).
\end{array}
\right.
\end{eqnarray}
\renewcommand{\arraystretch}{1}
Substituting (\ref{Eq:int1}) and (\ref{Eq:int2}) into (\ref{Eq:Cha}),
we find that
\begin{eqnarray}
C((\partial h_a)^2 e^{h_a}) &=& 
-8a Z(a) \left(\frac{1}{2}c_0-\frac{1}{4}(c_4+c_{-4})\right) \nn
&&
+8a \sum_{m=2}^\infty (-1)^m Z(a)^m
\left(
\frac{1}{4}(c_{2(m-1)}+c_{-2(m-1)})\right.\nn
&&
\hspace{1cm}
\left.
-\frac{1}{4}
(c_{2(m+1)}+c_{-2(m+1)})\right)\nn
&=& -4aZ(a) c_0 +2a Z(a)^2(c_2+c_{-2})\nn
&&+2a(1-Z(a)^2)\sum_{n=2}^\infty
(-1)^n Z(a)^{n-1}(c_{2n}+c_{-2n}).
\label{Eq:Qsecond}
\end{eqnarray}
Finally, from (\ref{Eq:Qfirst}) and (\ref{Eq:Qsecond}), we obtain
the oscillator expression of the modified BRS charge as in
(\ref{Eq:newBRS2}). 

\newpage
\section{Table of matter and ghost states at levels $\leq 6$}

The following table describe the matter and ghost states which
contribute to scalar fields at levels $\leq 6$. The inner products
$A_{ij}^{\rm mat}$ and $A_{ij}^{\rm gh}$ are defined by
eqs.~(\ref{Eq:Amat}) and (\ref{Eq:Agh}).

\begin{eqnarray}
\begin{array}{|l|r||l|r|}
\hline
 & \makebox[6cm]{\rm state}&
\makebox[5.5cm]{\rm inner products}\\
\hline
\hline
\ket{\eta_0} & \ket{0} & A^{\rm mat}_{0\,0}=1
\\
\ket{\eta_1} & (\alpha_{-1}\cdot\alpha_{-1})\ket{0}
&
A^{\rm mat}_{1\,1}=52
\\
\ket{\eta_2} & (\alpha_{-1}\cdot\alpha_{-2})\ket{0}
&
A^{\rm mat}_{2\,2}=52
\\
\ket{\eta_3} & (\alpha_{-1}\cdot\alpha_{-3})\ket{0}
&
A^{\rm mat}_{3\,3}=78
\\
\ket{\eta_4} & (\alpha_{-2}\cdot\alpha_{-2})\ket{0}
&
A^{\rm mat}_{4\,4}=208
\\
\ket{\eta_5} &
(\alpha_{-1}\cdot\alpha_{-1})(\alpha_{-1}\cdot\alpha_{-1})\ket{0}
&
A^{\rm mat}_{5\,5}=5824
\\
\ket{\eta_6} & (\alpha_{-1}\cdot\alpha_{-5})\ket{0}
&
A^{\rm mat}_{6\,6}=130
\\
\ket{\eta_7} & (\alpha_{-2}\cdot\alpha_{-4})\ket{0}
&
A^{\rm mat}_{7\,7}=208
\\
\ket{\eta_8} & (\alpha_{-3}\cdot\alpha_{-3})\ket{0}
&
A^{\rm mat}_{8\,8}=468
\\
\ket{\eta_9} & (\alpha_{-1}\cdot\alpha_{-3})(\alpha_{-1}\cdot\alpha_{-1})\ket{0}
&
A^{\rm mat}_{9\,9}=4368
\\
\ket{\eta_{10}} & 
(\alpha_{-2}\cdot\alpha_{-2})(\alpha_{-1}\cdot\alpha_{-1})\ket{0}
&
A^{\rm mat}_{10\,10}=10816,\ \ A^{\rm mat}_{10\,11}=416
\\
\ket{\eta_{11}} & 
(\alpha_{-1}\cdot\alpha_{-2})(\alpha_{-1}\cdot\alpha_{-2})\ket{0}
&
A^{\rm mat}_{11\,11}=5616,\ \ A^{\rm mat}_{11\,10}=416
\\
\ket{\eta_{12}} & 
(\alpha_{-1}\cdot\alpha_{-1})(\alpha_{-1}\cdot\alpha_{-1})
(\alpha_{-1}\cdot\alpha_{-1})\ket{0}
&
A^{\rm mat}_{12\,12}=1048320
\\
\hline
\ket{\chi_0} & \ket{1} 
&
A^{\rm gh}_{0,\,0}=1
\\
\ket{\chi_1} & b_{-1}c_{-1}\ket{1}
&
A^{\rm gh}_{1,\,1}=-1
\\
\ket{\chi_2} & b_{-1}c_{-2}\ket{1}
&
A^{\rm gh}_{2,\,3}=-1
\\
\ket{\chi_3} & b_{-2}c_{-1}\ket{1}
&
A^{\rm gh}_{3,\,2}=-1
\\
\ket{\chi_4} & b_{-1}c_{-3}\ket{1}
&
A^{\rm gh}_{4,\,6}=-1
\\
\ket{\chi_5} & b_{-2}c_{-2}\ket{1}
&
A^{\rm gh}_{5,\,5}=-1
\\
\ket{\chi_6} & b_{-3}c_{-1}\ket{1}
&
A^{\rm gh}_{6,\,4}=-1
\\
\ket{\chi_7} & b_{-1}c_{-5}\ket{1}
&
A^{\rm gh}_{7,\,11}=-1
\\
\ket{\chi_8} & b_{-2}c_{-4}\ket{1}
&
A^{\rm gh}_{8,\,10}=-1
\\
\ket{\chi_9} & b_{-3}c_{-3}\ket{1}
&
A^{\rm gh}_{9,\,9}=-1
\\
\ket{\chi_{10}} & b_{-4}c_{-2}\ket{1}
&
A^{\rm gh}_{10,\,8}=-1
\\
\ket{\chi_{11}} & b_{-5}c_{-1}\ket{1}
&
A^{\rm gh}_{11,\,7}=-1
\\
\ket{\chi_{12}} & b_{-2}b_{-1}c_{-2}c_{-1}\ket{1} 
&
A^{\rm gh}_{12,\,12}=1
\\
\hline
\end{array}\nonumber
\end{eqnarray}

\newpage
\section{Table of scalar states at levels $\leq 6$}

The following table lists scalar states at levels $\leq 6$.
$\left< \psi_i\right>_{(6,18)}$ denote vacuum expectation values of the
scalar states $\psi_i$ in level truncation calculations at $(6,18)$.

\begin{center}
\begin{tabular}{|l|r|r|r|r|r|r|r|}
\hline
 & & \multicolumn{6}{c|}{$\left<\psi_i\right>_{(6,18)}$} \\
\cline{3-8}
\multicolumn{1}{|c|}{\raisebox{1.5ex}[0pt]{$\psi_i$}} &
\multicolumn{1}{c|}{\raisebox{1.5ex}[0pt]{state}} &
\multicolumn{1}{c|}{\makebox[1.6cm]{$a=-0.50$}} &
\multicolumn{1}{c|}{\makebox[1.6cm]{$a=-0.40$}} &
\multicolumn{1}{c|}{\makebox[1.6cm]{$a=-0.25$}} &
\multicolumn{1}{c|}{\makebox[1.6cm]{$a=0.00$}} &
\multicolumn{1}{c|}{\makebox[1.6cm]{$a=0.50$}} &
\multicolumn{1}{c|}{\makebox[1.6cm]{$a=1.00$}} \\
\hline
\hline
$\psi_0$ & $\ket{\eta_0}\otimes\ket{\chi_0}$&0.00000
&0.49637& 0.56739& 0.54793& 0.48059& 0.42623
\\ \hline
$\psi_1$ & $\ket{\eta_1}\otimes\ket{\chi_0}$&0.00000
  &0.07446& 0.05809& 0.02857& 0.00038& $-$0.01221
\\ \hline
$\psi_2$ & $\ket{\eta_0}\otimes\ket{\chi_1}$&0.00000
  &$-$0.31083& $-$0.30955& $-$0.21181& $-$0.10466& $-$0.04878
\\ \hline
$\psi_3$ & $\ket{\eta_3}\otimes\ket{\chi_0}$&0.00000
  &0.01000& $-$0.00008& $-$0.00573& $-$0.00524& $-$0.00271
\\ \hline
$\psi_4$ & $\ket{\eta_4}\otimes\ket{\chi_0}$&0.00000
  &$-$0.00351& $-$0.00355& $-$0.00255& $-$0.00152& $-$0.00100
\\ \hline
$\psi_5$ & $\ket{\eta_5}\otimes\ket{\chi_0}$&0.00000
  &0.00426& 0.00176& $-$0.00016& $-$0.00055& $-$0.00018 
\\ \hline
$\psi_6$ & $\ket{\eta_1}\otimes\ket{\chi_1}$&0.00000
  &$-$0.02664& $-$0.01077& 0.00370& 0.00817& 0.00629
\\ \hline
$\psi_7$ & $\ket{\eta_0}\otimes\ket{\chi_4}$&0.00000
  &$-$0.05489& 0.00881& 0.05739& 0.06430& 0.04973 
\\ \hline
$\psi_8$ & $\ket{\eta_0}\otimes\ket{\chi_5}$&0.00000
  &0.04502& 0.04922& 0.03406& 0.01876& 0.01125 
\\ \hline
$\psi_9$ & $\ket{\eta_0}\otimes\ket{\chi_6}$&0.00000
  &$-$0.01830& 0.00294& 0.01913& 0.02143& 0.01658 
\\ \hline
$\psi_{10}$ & $\ket{\eta_6}\otimes\ket{\chi_0}$&0.00000
  &0.00273& 0.00178& 0.00175& 0.00180& 0.00138 
\\ \hline
$\psi_{11}$ & $\ket{\eta_7}\otimes\ket{\chi_0}$&0.00000
  &0.00049& 0.00128& 0.00146& 0.00108& 0.00075 
\\ \hline
$\psi_{12}$ & $\ket{\eta_8}\otimes\ket{\chi_0}$&0.00000
  &0.00118& 0.00084& 0.00072& 0.00066& 0.00049 
\\ \hline
$\psi_{13}$ & $\ket{\eta_9}\otimes\ket{\chi_0}$&0.00000
  &0.00111& 0.00025& 0.00014& 0.00036& 0.00031 
\\ \hline
$\psi_{14}$ & $\ket{\eta_{10}}\otimes\ket{\chi_0}$&0.00000
  &$-$0.00015& $-$0.00003& 0.00008& 0.00009& 0.00007 
\\ \hline
$\psi_{15}$ & $\ket{\eta_{11}}\otimes\ket{\chi_0}$&0.00000
  &0.00002& 0.00001& 0.00001& 0.00001& 0.00000
\\ \hline
$\psi_{16}$ & $\ket{\eta_{12}}\otimes\ket{\chi_0}$&0.00000
  &0.00012& 0.00003& 0.00000& 0.00001& 0.00001 
\\ \hline
$\psi_{17}$ & $\ket{\eta_3}\otimes\ket{\chi_0}$&0.00000
  &$-$0.00501& $-$0.00232& $-$0.00132& $-$0.00167& $-$0.0015 
\\ \hline
$\psi_{18}$ & $\ket{\eta_4}\otimes\ket{\chi_1}$&0.00000
  &$-$0.00059& $-$0.00074& $-$0.00058& $-$0.00034& $-$0.00022 
\\ \hline
$\psi_{19}$ & $\ket{\eta_5}\otimes\ket{\chi_1}$&0.00000
  &$-$0.00096& $-$0.00021& $-$0.00004& $-$0.00025& $-$0.00027 
\\ \hline
$\psi_{20}$ & $\ket{\eta_2}\otimes\ket{\chi_2}$&0.00000
  &$-$0.00012& $-$0.00008& $-$0.00006& $-$0.00005& $-$0.00003 
\\ \hline
$\psi_{21}$ & $\ket{\eta_2}\otimes\ket{\chi_3}$&0.00000
  &$-$0.00006& $-$0.00004& $-$0.00003& $-$0.00003& $-$0.00002 
\\ \hline
$\psi_{22}$ & $\ket{\eta_1}\otimes\ket{\chi_4}$&0.00000
  &$-$0.00506& $-$0.00131& $-$0.00168& $-$0.00376& $-$0.00371 
\\ \hline
$\psi_{23}$ & $\ket{\eta_1}\otimes\ket{\chi_5}$&0.00000
  &0.00126& $-$0.00038& $-$0.00142& $-$0.00123& $-$0.00084 
\\ \hline
$\psi_{24}$ & $\ket{\eta_1}\otimes\ket{\chi_6}$&0.00000
  &$-$0.00169& $-$0.00044& $-$0.00056& $-$0.00125& $-$0.00124 
\\ \hline
$\psi_{25}$ & $\ket{\eta_0}\otimes\ket{\chi_7}$&0.00000
  &$-$0.04011& $-$0.03204& $-$0.03001& $-$0.03071& $-$0.02579 
\\ \hline
$\psi_{26}$ & $\ket{\eta_0}\otimes\ket{\chi_8}$&0.00000
  &$-$0.00536& $-$0.01649& $-$0.01875& $-$0.01288& $-$0.00819 
\\ \hline
$\psi_{27}$ & $\ket{\eta_0}\otimes\ket{\chi_9}$&0.00000
  &$-$0.01448& $-$0.01167& $-$0.01142& $-$0.01206& $-$0.01018 
\\ \hline
$\psi_{28}$ & $\ket{\eta_0}\otimes\ket{\chi_{10}}$&0.00000
  &$-$0.00268& $-$0.00824& $-$0.00938& $-$0.00644& $-$0.00409 
\\ \hline
$\psi_{29}$ & $\ket{\eta_0}\otimes\ket{\chi_{11}}$&0.00000
  &$-$0.00802& $-$0.00641& $-$0.00600& $-$0.00614& $-$0.00516
\\ \hline
$\psi_{30}$ & $\ket{\eta_0}\otimes\ket{\chi_{12}}$&0.00000
  &$-$0.00765& $-$0.01029& $-$0.00773& $-$0.00413& $-$0.00243
\\ \hline
\end{tabular}
\end{center}


\end{document}